\pdfoutput=1 
\documentclass[10pt]{article}

\usepackage{amsmath}
\usepackage{amssymb}

\usepackage{graphicx}

\usepackage{cite}

\usepackage{color} 

\usepackage{wasysym}
\usepackage{pifont}
\usepackage[colorlinks]{hyperref}

\usepackage[labelfont=bf,labelsep=period,justification=raggedright]{caption}

\makeatletter
\renewcommand{\@biblabel}[1]{\quad#1.}
\makeatother

\date{}

\begin{document}

\begin{flushleft}
{\Large
\textbf{Rohlin Distance and the Evolution of Influenza A virus: Weak Attractors
and Precursors}
}
\\
Raffaella Burioni$^{1,\ast}$, 
Riccardo Scalco$^{2}$, 
Mario Casartelli$^{1}$
\\
\bf{1} Dipartimento di Fisica e INFN, Universit\`a di Parma, Parma, Italy
\\
\bf{2} Department of Biochemistry, University of Zurich, Zurich, Switzerland
\\
\bf{3} Dipartimento di Fisica e INFN, Universit\`a di Parma, Parma, Italy
\\
$\ast$ E-mail: Corresponding raffaella.burioni@fis.unipr.it
\end{flushleft}

\section*{Abstract}
The evolution of the hemagglutinin amino acids sequences of Influenza A virus  is studied by a
method based on an informational metrics, originally
introduced by Rohlin for partitions in abstract probability
spaces. This metrics does not require any previous functional or
syntactic knowledge about the sequences and it is sensitive to the
correlated variations in the characters disposition.  
Its efficiency is improved by algorithmic tools, designed to enhance the detection of the novelty 
and to reduce the noise of useless mutations.  We focus on the USA data from 1993/94 to
2010/2011 for A/H3N2 and on USA  data from 2006/07 to 2010/2011 for A/H1N1 . We show that the clusterization of the distance
matrix gives strong evidence to a structure of domains in
the sequence space, acting as weak attractors for the evolution,
in very good agreement with the epidemiological history of the
virus. The structure proves very robust with respect to the variations
of the clusterization parameters, and extremely coherent
when restricting the observation window. The results suggest an
efficient strategy in the vaccine forecast, based on the
presence of "precursors" (or "buds") populating the most recent
attractor.


\section*{Introduction}

There is a long history in approaching DNA and RNA sequences as texts, with
quantitative estimates for various kinds of statistical properties and
complexity indicators \cite {waterman,pevzner}. The general idea behind this
approach is that the information encoded in the sequence is strictly related to
the properties of the corresponding biological structures and that good
indicators should be able to recognize similar functions in different
sequences. 

The sector devoted to estimate the relevance of mutations along a time ordered
set of evolving sequences is particularly interesting when a sufficiently long
record of samples  is accessible, as for viral RNA of rapidly evolving diseases
\cite{hampson,history,Nelson,Liao}. We focus on a definite kind of statistical
properties, precisely metric properties. 

The distance between sequences is a concept admitting several implementations.
In the context of evolving viral RNA, distances based on the sequence symbols are mostly of the Hamming type:
for two strings $\mathbf{a}$ and $\mathbf{b}$ of characters, the Hamming distance 
$d_H(\mathbf{a},\mathbf{b}) $ is the number of sites with different symbols  \cite{ham,Deem,HeDeem}. Such distances are sensitive to
local features only, since mutations occurring at different sites are non correlated.  In this framework, an interesting solution has been proposed  in \cite{Deem,HeDeem}, where, focusing on particular locations of the sequence (the epitopes, whose role and peculiarities in Influenza Virus evolution are well known), authors succeeded in extracting important features of strains evolution. In a sense, the extra information introduced with the choice of the epitopes proved efficient in overcoming the intrinsic uncorrelation of the Hamming metrics, leading to interesting results. 
Other relevant approaches, based on sequences information only, rely on entropic distances
\cite{otusayood,Xia,ouyang,butte,rao} referred to the Shannon's entropy or to
compression algorithms, and they are mainly addressed to the comparison of
strings of different length in inhomogeneous frameworks, a procedure motivated
by the fact that in evolving sequences, beside substitutions, there are
frequent insertions and deletions \cite{bax}. 

There are few remarkable
alternatives to these "sequence based" type distances. Hemagglutination inhibition (HI) assays
\cite{emmaglutin}, reporting the ability of ferret antibodies, raised against
one viral strain, to inhibit a second strain's ability to agglutinate red blood
cells, are currently used to define similarity between antigens
\cite{Smith}.  Certainly, the metrics extracted from HI tests is  directly related
to the real antigenic similarity between strains, but it requires HI assay animal data, which are difficult to
obtain with high precision.

We intend to introduce an enhanced version of a different metrics, known as {\textit{ Rohlin distance}} \cite{AA,rohlin}, which is  based on the sequences symbols and is expected to be sensitive to their global distributions and correlations. 
It is also founded on the Shannon's entropy but, differently from other informational functionals,
applies in a biologically homogeneous framework. Moreover it does not deal with the frequency probabilities of symbols on single sites, which are too poor as units and do not touch the global structure (see however  \cite{Xia} for interesting
improvements in this directions). In our approach, the basic entities  are indeed  
the partitions of a sequence into subsequences, as they are determined, starting from a configuration (the list of amino acids), by a projection operation: precisely, we consider  the  {\textit{ partitions defined  by homogeneous segments}}. This aims at evidencing the ordered collection of connected subsequences of equal symbols. For instance, the alphabetical string $\mathbf{a} =AAABBACCCB $ would be
divided into five subsequences, and each subsequence would determine a segment. To the first segment $AAA$ there correspond the site subset with labels $(1,2,3)$, to $BB$ the subset $(4,5)$, etc. The natural length of each segment is the number of its symbols. In the example, the lengths are $3,2,1,3,1$. Once such lengths are correctly normalized, this assignment is equivalent to the definition of a probability measure on the subset algebra, proportional to the number of sites contained in each subset. Finally, partitions can be represented by their bounds, the segment extremes, or more economically by the left extremes, allowing for a simple and straightforward comparison between partitions.

In the partition space 
the Rohlin distance $d_R$ is then defined,  for any couple of partitions
$(\alpha,\beta)$, 
by the mutual conditional Shannon entropy (see Materials and Methods for details): 
$$
d_R(\alpha,\beta)=H(\alpha|\beta)+
H(\beta|\alpha)~.
$$
The conditional Shannon entropy $H(\alpha|\beta)$ represents the residual information needed to describe the segment disposition of $\alpha$ when the disposition of $\beta$ is known, or, in other terms, how the knowledge of $\beta$ may contribute to the knowledge of $\alpha$.  Therefore the symmetrized form above, defining $d_R(\alpha,\beta)$, is the total information required to distinguish $\alpha $ and $\beta$, seen as schemes of segments with their probabilities.

In absence of  a bias, our choice, assigning equal weight to each site and leading to a measure proportional to the length, is the most natural. Other probability measures can be defined, but they cannot depend on the configurations, since in that case the same set would have different measures for two configurations, and the conditional Shannon entropy would loose its meaning, inhibiting the very definition of Rohlin distance.

The segmentation provided by the partitioning of the sequences into homogeneous subsequences entails many advantages: its definition is simple and universal; sequences are not too tightly fractioned, as by single symbols; no {\textit{ a
priori}} knowledge is required, along with the exigence of a ``black box'' analysis.
Moreover, even if segments have no intrinsic biological meaning (and this could appear as an
inconvenience), alterations in their overall distributions, as those emerging in historical
records, are by definition compatible with the biologically efficient features proposed by
evolution \cite{Deem3}.

Thus, by using probabilities which arise from this geometrical and topological
asset,  the distance $d_R(\alpha,\beta)$ measures the information content carried by
evolution, giving evidence to the emerging dissimilarities. Clearly, this content should be
filtered, as far as possible, from the effects of the non evolving part: to this end we
introduce the so called {\textit{ reduction process $\pi$}}, a method designed to amplify the
relevant differences between partitions by dropping their common sub-partitions. This means that 
for any couple 
$ (\alpha,\beta)$ there is a reduced couple $(\widehat\alpha,\widehat\beta ) =
\pi (\alpha,\beta )$  
at amplified distance. From a practical point of view, the reduction consists
in erasing the common
extremes of segments between two partitions. A key point is that this method
proves surprisingly effective also 
in filtering  the noise of useless mutations. Details are given in the Material and Methods section. 

We shall deal here with an interesting example of a highly mutating sequence,
the RNA of influenza A virus, whose databases are particularly rich. More
precisely, we consider the amino acids sequence of the surface protein
hemagglutinin for H3N2 subtype Influenza A virus \cite{Lipman}, with human as
host, in $1470$ strains  collected in USA from 1993/94 to 2010/2011
\cite{Bao08}, and the analogous sequence for H1N1 subtype for $2506$ strains
collected in USA from 2006/07 to 2010/2011.   As our method works on equal length strings and  it is based on an informational metrics with long range correlations, it is expected to perform better when applied to the longest sequences. We therefore choose the full length (566) HA sequences, which are a subsample of the available sequences. We also consider sequences identified by a complete date, as the time when the sequence appears represents an important information in our analysis. The restriction to the USA sequences is motivated by several facts. First,  choosing sequences from  the temperate regions,  we give relevance to the seasonal timing of the virus evolution, minimizing the interference with a dephased development; second, the geographical bounds in the sampling ensure that we are looking, season by season, to a reasonably stable population. Notwithstanding these restrictions, for example for H3N2 we process 1470 sequences over 1986 in the northern hemisphere  (more than 74 $\%$)   and over 2759 in the world (more than 53 $\%$).  In other words, we keep the statistical majority, disregarding only possible noise. Analogous estimates hold for the processed H1N1 sequences.

After partitioning each sequence, we calculate the Rohlin distance between all the partition pairs, and
analyze the whole sequences sample by the hierarchical complete linkage
clustering algorithm \cite{Has09} on the distance matrix. The procedure has
strong analogies with the analysis presented in \cite{Levin,HeDeem} but in a
completely different metric space, namely the partition space instead of the configuration space.  Our analysis traces the evolution of
topological properties of the sequences, while the virus escapes following its
antigenic drift. Interestingly, the structures arising in the sequence space
according to this metrics result to be quite meaningful;  they individuate indeed 
well defined regions of the sequences space acting as {\textit{ weak attractors}} \cite{HeDeem,Smith},
where the evolution of the virus takes place for definite periods. Moreover,
the attractors display precursors \cite{HeDeem,Smith}, i.e. sequences populating the regions well
before they are identified as circulating strains, an information which appears
to be relevant in the forecast problem, suggesting an alternative
strategy in vaccines formulation. Once again there are analogies but also differences with the privileged regions of evolution presented in previous analysis \cite{HeDeem}. Our attractors arise indeed  from  of similitude related to the overall disposition of  homogeneous segments, and not to the actual data lying on every site (with the supplementary restriction to epitopes). This means that, more than an overlap of results, the two approaches present complementary points of view, enforcing one another. It is therefore quite remarkable that
the vaccine predictions of the two methods agree very well, a fact whose origin is not completely understood yet.

\section*{Results}

\subsection*{Clustering and weak attractors in Rohlin metrics}
Using the sequences set, we calculate the whole $N\times N$ matrices of Hamming
and Rohlin distances, 
i.e. ${\mathbf H}_{ik} $,  ${\mathbf R}_{ik} $ and reduced Rohlin distance
$\widehat{\mathbf R}_{ik} $, whose entries are respectively $d_H(\mathbf{a}_i ,
\mathbf{a}_k) $ on sequences, or $d_R(\alpha_i,\alpha_k) $ and
$d_R(\widehat\alpha_i,\widehat\alpha_k) $  on the corresponding partitions
(possibly simplified by the reduction process). The primitive (non-reduced)
partitions result to be by far less informative than the reduced ones, so that
from now on we shall omit to report  calculations on them. 

This fact is a non trivial result in itself:
in principle, by deleting the common bounds, the reduction process could imply indeed the survival of a random set of unstable bounds created by the incoming mutations. This happens, as expected, with artificially created random mutations (as reported in the last subsection). Clearly, far from being more informative, such chaotic residual set would be useless for our scope.

The empirical fact we observe is quite different: once the reduction process has cleaned the useless common sub-partitions,  the surviving bounds (defining the new reduced partitions), far from being ``random'',  carry the winning novelty in the adaptive strategy of the virus. In other terms, the effective new disposition of segments is not random because it has been selected. Moreover, the mutual disposition has intrinsically a long range character, which is captured by such a non local metrics as the Rohlin distance is (details on the amplification are given in the mathematical section in Materials and Methods).

The scenario emerges from a ``clustering analysis''  on the matrix
$\widehat{\mathbf R}_{ik} $. The used clustering tool is the standard
hierarchical complete linkage algorithm \cite{Has09}, where the number $p$ of
clusters is an external  parameter. An important point is therefore the choice
of the optimal $p$.  
Since at fixed $p$ every clusterization corresponds to a partition of the whole
set into groups populated by $n_1, n_2,...,n_p$ sequences, with $n_1+
n_2+...+n_p=N$, by defining the probabilities $w_i = n_i/N,~i=1,2,...,p$  to each
clusterization we can associate the Shannon entropy  $H(w_1,...,w_p) = -\sum_i
w_i \log w_i$ of its probability distribution. Since at growing $p$ the cluster
populations cannot increase, the entropy is non-decreasing  \cite{khinchin}.
However, if it is substantially stable, this means that the clusters are also
stable (or equivalently, that newly added probabilities are very small); if it
grows, then clusters (and probabilities) are almost continuously splitted into
smaller and smaller ones. Now, observing the behaviors of Rohlin and Hamming
entropies in Figure 1 for H3N2, we note a clear quasi-plateau for Rohlin in the
interval $14 < p < 34$, while Hamming is always growing. For the latter,
the interpretation is that the Hamming clusters, being an artificial product of
the procedure, split with a remarkable continuity. 

As to Rohlin, the long
plateau  clearly indicates that the clusters are real structures in the
sequences space, keeping a definite individuality in a large observation range. 
The growth for low $p$ is simply due to the fact that, if the imposed number is
too small, ($p=1,2,3...$), the calculated clusters must contain the real ones,
so that at growing $p$ the splitting is effective, up to the optimal number
when calculated and real clusters coincide. On the contrary, for $p$ too large,
calculated clusters are so numerous that 
also the real ones begin to split much more effectively than during the
peripheral loss registered in the plateau.
The plateau extremes may be therefore roughly related to the typical isolation
length among real clusters, and to their maximal diameter respectively.
Interestingly, the optimal value {$p=14$, obtained from clustering without
additional hypotheses, is consistent with  the number of different circulating
strains identified by the WHO  HI tests. 

Analogous results obtained for H1N1
are shown in Figure S1. In that case, the optimal value for the clustering
parameter is $p=8$.

Since every sequence is marked by its sampling date, a natural question is the
time distribution of the resulting clusters. The upshots for H3N2
are summarized in Figure 2.  The part above refers to the
clustering on $\widehat{\mathbf R}_{ik} $. Below, for comparison, to the
clustering on  ${\mathbf H}_{ik}$. There is no scale on the $Y$ axis because
the ordinate only distinguishes among clusters (same ordinate $\equiv$ same
cluster): order and color  have no intrinsic meaning and they have been chosen
with readability criteria only. The $11$ different polygonal symbols represent
the $11$ reference viruses, including the alternates, observed in the Northern
Hemisphere according to WHO HI tests. Their names are indicated in the
plot and the details of the sequences are given in Table 1. The reference sequences are
not identified by a time $X$ coordinate. They are processed together with the dataset and they 
are positioned in the clusters they belong to after the Rohlin clustering procedure. Vertical lines separate winter seasons and are conventionally set at July 31. 

From the $d_R$ clustering of Figure 2, 
which is very stable by changing $p$ in the plateau range of Figure 1,  
we can draw several indications. First,
there is a clear long temporal extension of clusters, which are densely
populated for several winter seasons. Interestingly, they present precursors,
that we term "buds", and successors, i.e. a bunch of sequences representing viruses that appear, in time, before
or after the main part of the cluster \cite{HeDeem}. The identification of buds will be explained in details in the next sections.

Let us consider, for instance, the $94/95$ season. In that year, we observe a bud (Wuhan strain, 
the yellow cluster)  which is getting stronger in the following season, living jointly with the 
Johannesburg strain, the red cluster. Then, it becomes the dominant strain in 
$95/96$ and $96/97$, while it may be considered a successor in $97/98$, when the Sidney (light green)
is the dominant strain. Notice that the same Sidney was a bud  in $96/97$. 

The distribution of clusters suggests that the evolution in the
sequence space takes place in preferential regions, corresponding to each
cluster, which can be populated well before and after the main season. Such
regions act as a kind of weak (i.e. non definitive) attractors. For example, as mentioned,
two H3N2 virus strains circulated during 96/97 winter season: Wuhan successors
and Sydney buds. Through HI tests, WHO revealed the Wuhan reference virus as
the circulating one, recommending it as a vaccine for the season 97/98. It is
crucial that, already during 96/97, our analysis shows the emergence of a bud,
the Sydney family strain, which is the actual virus circulated in 97/98 winter
season. Rohlin attractors correctly describe also the heterogeneity coming from
"outliers" sequenced by the WHO, which  must not be treated separately, as it
happened in other Hamming based approaches \cite{Levin}. These sequences
naturally fall into a cluster, confirming that Rohlin correctly takes into
account the variability present in outliers.  

A second point is that $d_R$-clustering is consistent with epidemiological
WHO-HI data. For example, the subsequent strains A/Wuhan/359/95, A/Sydney/5/97
and A/Moscow/10/99, appeared during years from 95/96 to 02/03 according HI
tests, are represented by three well defined clusters (A/Panama/2007/1999 is 
a Moscow alternate). Interestingly,  their
reference sequences belong to the correct clusters once they are included in
the data set, without any a priori information. 

In the lower part of Figure 2 the same clustering procedure is shown referring 
to the Hamming matrix ${\mathbf H}_{ik}$. A definite temporal extension of clusters 
is observed, in agreement
with previous results \cite{Levin,HeDeem}. However, the cluster temporal distribution
obtained from ${\mathbf H}_{ik} $ is quite unstable, confirming the dependence
on $p$ evidenced in Figure 1. The $p$ used here is  the same of Rohlin, but the choice is 
completely arbitrary because there is not a clear plateau for Hamming.
This means that some appearing spots are not true  buds, as they results from the almost
continuous splitting of Hamming clusters, which are not stable under a change in
$p$. Namely, a new cluster can be produced simply by raising $p$. 
Moreover in some seasons (e.g. from 99/00 to 02/03)
there is a contrast between WHO indication and the cluster arrangement:
some clusters are not represented by any reference sequence, while others are
wrongly doubly represented, showing a poorer correlation with HI analysis.  For example, Sydney and Moscow reference strains belong to the same cluster while they are expected to be in different ones.

\subsection*{Buds in Rohlin weak attractors and vaccine forecast}
An interesting evidence can be drawn from the position of symbols in the
attractors.
The symbols above the upper diagram in Figure 2 show in the first row the vaccine
indicated by the WHO on the basis of the HI tests in previous seasons, while in
the second row they represent the indication we would suggest on the basis of
the following criterion: by looking at the $d_R$-cluster distribution, when in
a year there are simultaneous strains (with statistically significant populations), 
we would indicate the newest one, i.e.
the bud, as emergent in the next year. This criterion, in other words, sees the
novelty carried in emergent buds as a feature enhancing the aggressiveness of
the virus. In both the WHO or buds analysis, the symbols are shown in green
when the vaccine choice agrees with the circulating strain, in red  when they
disagree.We used a two-colors symbol when more than one virus circulated in
the same season and the corresponding prediction agrees with one of the two circulating strain. 
Now,  the second symbols row indicates that the buds criterion is able to identify correctly
the circulating strain every year, apart from the season $01/02$ because of the lack of sequences, 
while the WHO criterion fails in  3  cases over 17. Of course, it will be extremely interesting to verify the criterion
on the set of sequences for next season, as soon as they will be available.

In the lower part of Figure 2, displaying  the results of the same procedure on the Hamming
distance, we note that buds as early warning of new strains are not reliable because, as discussed above, the instability in $p$ does not allow for an unambiguous detection of their appearance, as they can be produced by raising the value of $p$.

The bud criterion can be successfully applied also to the very interesting case
of H1N1 and the results are shown in Figure 3. 
The sampling for H1N1 is very inhomogeneous. From 2006 the statistics
increases, so we will limit our analysis to this period.  From the clustering
entropy analysis (shown in Figure S1), the optimal  $p$ is   
$8$. The very relevant cluster starting suddenly around April 2009 (red
line in Figure 3) \cite{Morens} represents precisely the pandemic virus appeared
in the $2009$ season. In that case, the bud criterion partially fails, as it
recognizes correctly only the strains circulated at the beginning of the
season. This is reasonable, since our method is expected to be effective in
simple antigenic drift, and not in the case of a dramatic change, as in the
$2009$ pandemic case. The shift probably sets a completely new ``direction" in
the sequence space. Notice that our method evidences the simultaneous
occurrence of four well distinct clusters in 2009, a  feature missing in the
Hamming analysis. The observation of multiple clusters signals could be related
to a typical instability of post-pandemic  periods, as the one we are facing
according to the WHO
\cite{Morens}. We expect this structure to be present also in the new set of
sequences for season 2011/12.

\subsection*{Restricting the time window}
Another natural question, the relevance of the examined time window, is treated
in Figure 4, displaying the results one would  obtain by stopping the data
collection at five different years, i.e by applying the clustering to the
restricted sets of sequences available at those times. This procedure is
intended to clarify how the bud criterion works and to check that it is not an unpredictive {\textit{ a
posteriori}} verification of the vaccine choice, but a real working framework.

The time distribution in the $X$ axis describes at various years the position of the H3N2  sequences, exactly
as in the upper Figure 2. However, to reproduce exactly the 
situation in which the WHO vaccine prediction is made, we collect all the sequences 
available up to March
of a given year and we perform our clustering analysis only on that dataset.  There is
therefore an increasing number of sampled sequences, in every horizontal time sector starting for the
upper part down to the lowest.
In details,  we apply our entropy criterion each time to the dataset restricted to the end of the winter season of a given year and, from that dataset only, we choose the optimal $p$ for that case. Interestingly, there is a clear plateau in the entropy analysis for every restricted time window, allowing for a unambiguous choice of the number of clusters. The Rohlin entropy analysis from the restricted time window is shown in Figure S2, together with the corresponding one for Hamming. Notice that for the Hamming clustering, there is no clear indication of the number of clusters $p$ from the entropy analysis, nor any evidence of emergent buds. 

In principle, leaving out part of the data, these
clusterings could be different from the final one. It is remarkable that the 
 structure remains the same. Buds are clearly present, as if the evolution
 took place in a well defined landscape, with preferential ``antigenic"
 directions that are filled during the genetic drift, and acting therefore as
 weak attractors. The symbols of the reference WHO strains are excluded from  the clustering in the time restricted window, as they would represent an a posteriori knowledge. They are associated to 
each cluster by an inverse analysis, i.e. by calculating the reference WHO strain 
which has minimum distance with the sequences belonging to the bud cluster. Details on the reverse analysis, applied also to the
whole dataset are given in the next subsection. Now, in Figure 4 a syringe indicates the most accredited
 vaccine for the next year based on the bud criterion: even when the previous
 years database is poor, the forecast is very good and the cluster
 corresponding to the syringe is exactly the prevailing cluster observed the
 next year. By confirming the coherence of the procedure, this result supports
 the bud emergence criterion for the prediction of the new prevailing strain. 

Interestingly, our bud criterion, which does not include any additional information on epitopes positions, agrees very well with the dominant strain prediction discussed in  \cite{HeDeem}.  This appears to indicates that  the "black box" Rohlin distance analysis is able to grasp the biological information included in the "epitopes" metrics, which is certainly correct but requires an additional input (the relevant positions). We notice that our results for the vaccine choice always agree with the prediction of  \cite{HeDeem} when there is a single circulating strain, while they are complementary when there are two. We do not have a clear explanation for this interesting fact, at the moment. 

 \subsection*{Testing the method: reverse analysis and random permutations}

The clustering procedure can be performed by a completely different method,
which does not consider distances between sequences themselves, as in the
hierarchical method, but refers to the WHO different sequences identified by
the HI analysis. Precisely, the Rohlin distance has been calculated between
each of the 1470 sequences of the H3N2 with all the $11$ WHO reference virus strains.
The sequences are temporally aligned along the $X$ axis of Figure S3,
 while the reference strains have a conventional position
along the $Y$ axis. In the diagram, each sequence is represented 
by a point  whose $X$ coordinate is
its sampling date, and whose $Y$ coordinate corresponds  to the {\textit{nearest}} 
WHO reference strain. Surprisingly, the final result of
this new procedure is almost the same as the one showed in Figure 2 of the main
text, while it is obtained from a completely different analysis. The analogous
plot  on Hamming distance does not preserve the cluster structure.  Once again,
the Rohlin distance approach proves robust and consistent with the HI tests
analysis.
  
 As a further check of the robustness of our results, we consider a random permutation of the site labels,
 simultaneously performed on all the sequences of the H3N2 dataset. This operation leaves the
 Hamming distances invariant by definition. Since such a random mixing of the
 amino acids is biologically meaningless, a natural request is that the
 conclusions drawn from a correct metrics should crash: this is precisely what
 happens with the Rohlin distance.
 In other words, with Rohlin, only the partitions corresponding to the real
 sequences seem to encode correctly  the antigenic drift during the evolution,
 evidencing  a meaningful relation with the global structure of the sequences.
 Vice versa, the simple global ``mutations counting'' completely fails to recognize
 the information deletion caused by the label permutation. The results are
 presented in Figure S4.
 
\section*{Discussion}
The mechanism underlying influenza A antigenic plasticity, that is, how the
virus continually escapes the immune system by producing variant strains that cause
re-infection within a few years, remains an outstanding evolutionary problem.

There are two main general pictures for this evolution. The first one is based
on an almost continuous slow drift from an ancestor sequence, with some large
shifts occurring at certain stages of the evolution \cite{Fitch}. The second
one relies on a punctuated evolution, where "antigenic" swarms of sequences
populate ``basins'' in the sequence space for several years,  until the
circulating swarm ``jumps'' to another basin, reinfecting the population.
Dynamical model for this type of evolution have been built \cite{Koelle}, and
the evidence of a punctuated antigenic evolution has been put forward by
several authors \cite{Smith,Levin,pybus,Recker,holmes}. 

The picture emerging from the Rohlin metrics seems to support such a punctuated evolution with a
better fit of epidemiological data, giving also insights on the relevant
distance between circulating strains and vaccines. In fact, clusters result to
be organized into well defined regions of the sequences space: the virus
appears to explore for several seasons the sequence space region corresponding
to a certain  Rohlin width, until a jump takes it to another attractor, where
the evolution  starts again (a return to a previous region is also possible).
These regions constitute therefore {\textit{ weak attractors}} in the sense that they
are able to trap the virus for a finite time, and they can also be re-populated
after years. In other words, weak attractors seem to identify privileged
antigenic directions from genetic data. 

Interestingly, the clusters present
``precursors'', that we termed  {\textit{ buds}}, i.e. a small number of sequences
which explore in advance the next attractor when most of the strains still
belong to the previous one. All goes as if such precursors manage to experience
the winning escape strategy, that will be followed by the main swarm in
subsequent years, and a clear correlation emerges between the bud, i.e. the
younger attractor appeared in a given year, and the circulating strain of the
subsequent season. This {\textit{ bud criterion}}, in parallel with HI analysis,
could be helpful in the correct choice of the vaccines. The picture emerging
from Rohlin distance analysis appears to hold also by processing analogous data
sets as the A/H1N1 in USA. Interestingly, in H1N1 the bud criterion
partially fails in 2009, as it  recognizes correctly the emerging bud only
before the pandemic period, while it is not able to predict the clear new
cluster that appears suddenly in April 2009. The analysis correctly signals
also the high instability of the post-pandemic phase in $09/10$.

In conclusion, some main points should be stressed: the first is that no {\textit {a
priori}} knowledge of biological nature has been used or put into the data set.
The indications we have derived from clustering on the distance matrix
constitute a genuine emergence. It seems plausible therefore that the same
approach could work in similar circumstances, i.e. when a homogeneous set of
equal length arrays are at disposal. The second point is the existence of
structures in the sequence space, that can be described as weak attractors,
where the evolution of the viral species takes place with a discontinuous
dynamics. Clearly, a clustering algorithm is expected to recognize a
chronological order within the distance matrix, whenever the distance is a
monotone function of time, but in that case one would also expect the
progressive fragmentation of clusters as the external parameter $p$ grows. Such
is, substantially, the behavior suggested by the analysis on the Hamming
matrix. On the contrary, the presence of precursors \cite{HeDeem}, which discontinuously
anticipate the onset of future attractors, and the stability of the attractors
structure at varying $p$ or sampling, are quite non trivial facts, implying
that the Rohlin attractors are not a conventional decomposition in the sequence
space; they possess instead a robust, intrinsic, ``natural'' meaning. It seems
therefore that the Rohlin distance on reduced couples is able to evidence a
selected variety of admissible  ``antigenic states'', preferentially explored
through mutations, which remains hidden in other metric approaches. The third
point is that the ``buds emergence criterion''  could offer a valuable
complementary tool for an optimal strategy in the choice of vaccines. The
matter is obviously delicate, and a long series of experimental checks should
explore and confirm such a possibility before practical utilization, but we
think that an effort in this direction is worthwhile.

\section*{Materials and Methods}

\subsection*{Data}
The main
database
of reference for H3N2 is constituted by 1470 full-length (566 aa) HA proteins of the H3N2
subtype Influenza A virus isolated in USA  from 1993/94 to 2010/11,
excluding sequences with an incomplete sampling date. Such set is 
enriched with the $11$ reference sequences corresponding to the reference
viruses circulated in the same years according to WHO HI analysis \cite{fludb}.
The total is $N=1481$ sequences , written in the 20
amino acids alphabet.  As for H1N1, the main
database
of reference is constituted by $N=2506$ full-length (566 aa) HA proteins of the H1N1 subtype Influenza
A virus isolated in USA  from 2006/07 to 2010/2011, excluding sequences with
an incomplete date. Such set is  enriched with the $4$ reference
sequences, corresponding to the reference viruses circulated in the same years
according to WHO HI analysis \cite{fludb}.
The total is $N=2510$ sequences, written in the 20
amino acids alphabet.

\subsection*{Sequences and Rohlin Metrics}
Let $\mathbb{K}$ be a finite alphabet of characters (amino acids in our case). A sequence 
$ \mathbf{a} \equiv (a_1,a_2,...,a_L) $, where $a_k \in \mathbb{K}$, 
 may be thought as a function  on the one dimensional array $\mathbf{M}$ 
of sites labeled $1,2,...,L$. This function defines the {\textit {configuration}} or {\textit{ state}} 
on $\mathbf{M}$. Every sequence $ \mathbf{a}$ is therefore an element in a 
configuration space ${\mathcal{C}(\mathbf{M})}$. A probability measure $\mu$ on the finite subset algebra $\mathcal{M}$ of $\mathbf{M}$ is given by the normalized number of sites in every subset. This means that all sites are assumed to be equivalent. For instance, if $L=12$ and a subset $E$ includes sites $(2,5,10)$, then $\mu(E)=0.25$. 
Other measures are possible, e.g. by assigning weights to the sites. However, such weights (and measures) cannot depend on the configurations, otherwise the same subset of $\mathbf{M}$ could have more than one measure simultaneously, and the functionals defined below would loose any meaning.

A partition $\alpha \equiv ( A_1, ...,A_m  )$ of $\mathbf{M}$ is an exhaustive  collection of disjoint subsets
(called ``atoms'') of $\mathbf{M}$. The space $\mathcal{Z} (\mathbf{M}) $ is the set of all possible partitions of
$\mathbf{M}  $, where a partial order $\alpha \leq \beta$ means that $\beta$ refines $\alpha$.
The ``product'' $\gamma = \alpha \vee  \beta \equiv \alpha\beta$ (a close analogous of the minimal common multiple) is the minimal partition refining both factors $\alpha$ and $\beta$. The unit partition $\nu$ has only one atom, the whole set $\mathbf{M}$. Obvious properties such as $ \nu \leq \alpha = \alpha\alpha \leq \alpha \beta$ 
easily follow for every $\alpha$ and $\beta$.

The operation $\sigma  = \alpha \wedge  \beta $ is the maximal common factor (i.e. the most refined common sub-partition) of $\alpha$ and $\beta$. Clearly, $\nu \leq \alpha \wedge \beta \leq \alpha$, etc.

Every partition may be thought as an experiment where an elementary or {\textit{ atomic}} event $A_k $ occurs with probability $\mu(A_k)$.
Then, the meaning of the definitions above is that $\nu$ is the trivial experiment, $\alpha \leq \beta$ means that $\alpha$ is a sub-experiment of $\beta$, etc. 

 The Shannon's entropy $H(\alpha)$  is  defined as:
\begin{equation}
\label{shannon1}
  H(\alpha)= -\sum_{i=1}^m \mu(A_i)\ln \mu(A_i) ~.
\end{equation}
 If $\beta=(B_1,...,B_s)$, is another partition, the conditional entropy $H(\alpha|\beta)$  is 
\begin{equation}
\label{shancond}
  H(\alpha|\beta) = -\sum_{i=1}^m\sum_{k=1}^s \mu(A_i\cap B_k)\ln\frac{\mu(A_i\cap
  B_k)}{\mu(B_k)}~= ~ H(\alpha\beta)- H(\beta). 
\end{equation}
These quantities give respectively the mean incertitude of an experiment $\alpha$ and the residual mean incertitude on $\alpha$ when  the result of $\beta$ is known \cite{khinchin}. 
Now, for all $\alpha $ and $\beta  $ in $ \mathcal{Z} (\mathbf{M}) $, the Rohlin distance  is:
\begin{equation}
\label{rohlin1} d_R (\alpha,\beta)
=H(\alpha|\beta)+H(\beta|\alpha)~.
\end{equation}
The useful formula 
\begin{equation}
\label{rohlinbis}
d_R (\alpha,\beta)
=2H(\alpha\beta)-H(\alpha)-H(\beta)
\end{equation} 
follows from  Eq. \ref{shancond}.
Thus, $d_R(\alpha ,\beta ) $ is a measure of the overall non-similarity between  $\alpha $ and $ \beta $, giving account of 
the mutual correlations among the respective outcomes \cite{AA,rohlin}.  These concepts and definitions hold true in all probability spaces. For discrete spaces (graphs or lattices), where the {\textit{ states}} or {\textit{ configurations}} are determined by the values assumed by sites $j$'s in a finite alphabet  $\mathbb{K}$, $ d_R$  is therefore deeply different
from the well known Hamming distance $ d_H$ between configurations $\mathbf{a}$ and $\mathbf{b}$. This distance is defined,
up to a possible normalization factor, by: 
\begin{equation}\label{hamming}
d_H ({\mathbf{a}},{\mathbf{b}}) = \sum_{j}
|~b_{j}-a_{j}~|~.
\end{equation}
where $~|~x-y~|~$ is a distance in $\mathbb{K}$  if the alphabet is numerical, otherwise is 
1 for $a_j \not= b_j $  and 0 for $a_j = b_j $ (as in our case, since $\mathbb{K}$ is the alphabet of amino acids).
Counting the sites with different symbols regardless of their position, $d_H$ tells one nothing about  correlations between mutations. It is important to stress that the Hamming and Rohlin distances are
not defined on the same objects, the former being between configurations in ${\mathcal{C}(\mathbf{M})}$,
the latter between partitions in $  \mathcal{Z} (\mathbf{M}) $. 

In our particular case, where $\mathbf{M}$ is a one-dimensional finite lattice, and  the states (or configurations) are character sequences of length $L$, } we shall work with partitions generated by {\textit{ homogeneous segments}}, 
i.e. consecutive sites with the same value in  $\mathbb{K}$. Of course, in ${\mathcal{Z}(\mathbf{M})}$ 
there exist much more partitions, e.g. those with non connected atoms. 
As an example with $L = 12$, consider the fictional configuration $ \mathbf{a}=\{ A~A~A~T~T~C~A~A~A~F~F~B~\}$
The atoms (indicated by the site labels) of the corresponding partition $\alpha = \Phi (\mathbf{a})$ are  
$ \{ (1,2,3),(4,5),(6),(7,8,9),(10,11),(12) \}  $. The map $\Phi: \mathcal{C}  \to \mathcal{Z}$ is univocal but non invertible, since several  configurations are mapped into the same partition. For instance, a mutation from $C$ to $B$ 
as in $ \{...~T~T~C~A~A~...\} \to \{...~T~T~B~A~A~...\} $, does not affect the boundaries,  and it leaves the segment structure unchanged. Thus, by  the correspondence $ \Phi  $ and the Rohlin distance, we can evaluate ``how different'' are the states on $\mathbf{M}$ with regard to the correlated  distribution of segments. It is true that there is a  
loss of information due to the projection of many configurations into the same partition; but a comparable loss takes place 
also for Hamming, since the single site contribute gives account only for the ``equal-or-not'' distinction in $\mathbb{K}$. Moreover, as noticed, sites in $d_H$ are always totally uncorrelated.

The non-similarity between two partitions could be confused and weakened by the presence of a tight common factor, that we would eliminate as far as possible, in order to  amplify the Rohlin distance giving evidence to the real emerging novelty. However, such a ``reduction'' operation (analogous to the reduction to minimal terms for fractions) is not uniquely defined because partitions do not admit a unique factorization into primes \cite{reduc,reduc2}. The role of prime (i.e. indecomposable) factors can be played by 
{\textit{ dichotomic}} sub-partitions, which are still extremely redundant. Then, the key point consists in defining for each partition a restricted family $\mathbf{E} (\alpha ) $ of ``elementary'' dichotomic factors, with the following features:
\begin{enumerate}
\item  $\mathbf{E} (\alpha ) $ must be well defined for every  $ \alpha \in \mathcal{Z} $, or at least 
in the subset of $\mathcal{Z}$ actually under investigation;
\item   $\mathbf{E} (\alpha ) $ does not contain more factors than the number $m$ of atoms in $\alpha $;
\item  $\vee_{k=1}^m   \widetilde {\alpha}_k = \alpha $. 
\end{enumerate}

Now, assuming that the elementary factors  families $\mathbf{E} (\alpha ) $ and $\mathbf{E} (\beta ) $ have been  
defined, the reduction process consists in the following steps: 
\begin{enumerate}
\item  define the maximal common divisor $\sigma = \alpha \wedge \beta $;
\item  drop from   $\mathbf{E} (\alpha ) $ and $\mathbf{E} (\beta ) $ those factors which
are not relatively prime with $\sigma $, and note the surviving factors 
$ \widehat \alpha _k $ and $ \widehat \beta _j$ respectively 
(i.e. $ \widehat \alpha _k \wedge \sigma  = \widehat \beta _j \wedge \sigma  = \nu $);
\item  define $ \widehat \alpha  = \vee _k \widehat \alpha _k$ and  $ \widehat \beta  = \vee _j \widehat \beta _j$.
\end{enumerate}
In other words, we drop those dichotomic factors which are subfactors of the maximal
common factor $\sigma $, and the reduced $(\widehat \alpha ,\widehat \beta )$ are generated by the surviving factors.
 The amplification of non-similarity is a consequence of the following property:

\noindent Proposition:  $d_R(\widehat \alpha ,\widehat \beta )\geq  d_R(\alpha ,\beta ) $. 

\noindent The proof is elementary recalling that, if $\sigma = \alpha \wedge \beta$, we can write $\alpha = \sigma \widehat{\alpha}$ and $\beta = \sigma \widehat{\beta}$: indeed, as mentioned, $\sigma$ contains all the factors dropped during the reduction. Therefore, by formula (\ref{rohlinbis}) and the fact that $\sigma\sigma =\sigma$, the thesis
$ d_R(\alpha,\beta) \leq d_R(\widehat{\alpha} , \widehat{\beta})$  can be rephrased as
\begin{equation}
2H(\alpha \beta) - H(\alpha)- H(\beta) \equiv 2 H(\sigma\widehat{\alpha}\widehat{\beta}) 
- H(\sigma\widehat{\alpha}) - H(\sigma \widehat{\beta}) 
\leq 2 H (\widehat{\alpha}\widehat{\beta}) - H(\widehat{\alpha})- H(\widehat{\beta})
\end{equation}
Moving terms between the sides
\begin{equation}
2 H(\sigma\widehat{\alpha}\widehat{\beta}) - 2 H (\widehat{\alpha}\widehat{\beta})
\leq H (\sigma\widehat{\alpha}) -    H(\widehat{\alpha}) + H(\sigma \widehat{\beta}) 
- H(\widehat{\beta})~,
\end{equation}
and using formula (\ref{shancond}) for the conditional entropy, the thesis reduces to
\begin{equation}
2H(\sigma |\widehat{\alpha}\widehat{\beta}) \leq H(\sigma|\widehat{\alpha}) +
H(\sigma | \widehat{\beta})~.
\end{equation}
But this is clearly true since 
$$ H(\sigma |\widehat{\alpha}\widehat{\beta}) \leq H(\sigma|\widehat{\alpha}) ~~ {\mathrm {and}} ~~ H(\sigma |\widehat{\alpha}\widehat{\beta}) \leq H(\sigma |\widehat{\beta})~, $$ 
because the conditioning terms are greater in the left sides, q.e.d.

It is important to remark that this amplification regards the anti-similarity of the couple as a whole, while for reduced partitions as single entities the complexity possibly decreases, as expected: this means 
$H(  \widehat\alpha) \leq H(\alpha)$, etc. 

The correspondence $\pi ~:~(\alpha ,\beta ) \rightarrow  (\widehat \alpha ,\widehat \beta )$ defining this 
reduction process is many-to-one and idempotent. It is a projection from $\mathcal{Z} \times  \mathcal{Z}$ 
on the subset of irreducible pairs. The  process, therefore, essentially depends on the family 
of elementary factors, a choice which {\textit{ a priori}} can be implemented in many ways, reflecting the kind 
of interest the observer has in the experiment. Details and  procedures in abstract probability spaces may be found in \cite{reduc,reduc2}. Here we sketch an algorithmically easy recipe, fitting the very special case of character strings.

By exploiting one-dimensionality, a partition into segments (connected subsequences) can be economically represented by 
the list of the left bounds of segments. In the example above, $\alpha $ is fully determined by $(1,4,6,7,10,12)$. 

This suggest a very convenient choice of the family of elementary factors: precisely, for every $\alpha = \{A_1,A_2,...,A_k,...A_N \}$, the $k$-th dichotomic factor
$\widetilde{\alpha}_k \in \mathbf{E} (\alpha )$ is  
$$\widetilde{\alpha}_k = \{ \bigcup_{j=1}^kA_j, \bigcup_{j=k+1}^NA_j \}~,$$
and therefore, in terms of labels, the example above for $\alpha$ gives  $\widetilde{\alpha}_1 = (1,4) $, $\widetilde{\alpha}_2 = (1,6) $, etc.
With such a choice, the reduction process $ \pi $ described above consists in erasing 
all the common labels apart the first one (label 1 is indeed the necessarily common bound for alignment). For instance, consider again $\mathbf{a} $ as above, and a new configuration 
$\mathbf{b} = \{ A~A~H~H~H~Q~Q~Q~Q~A~Q~Q \} ~.$ 
The list for $\alpha $ is $(1,4,6,7,10,12)$, the list for $\beta  = \Phi ( \mathbf{b})$
is $(1,3,6,10,11) $, the list for $\sigma = \alpha \wedge \beta$ is $(1,6,10)$. Then, the reduced $\widehat \alpha  $ and $ \widehat \beta $ are represented by  $(1,4,7,12)$ and $(1,3,11)$ respectively. Note that they 
do not correspond to any new sequences, since the reduction is performed directly in  $ \mathcal{Z} (\mathbf{M})$, not in $ \mathcal{C} (\mathbf{M})$. A graphic intuitive representation of this reduction is given in Figure 5.

\section*{Acknowledgments}
This work is partially supported by the MIUR FIRB grant RBFR08EKEV and by the INFN project  "Biological applications of theoretical physics methods". The authors thank Roberto Burioni and Nicola Clementi for useful discussion and suggestions.
\bibliography{rohlinplos.bib}

\newpage
\section*{Figure Legends}
\begin{figure}[!ht]
\begin{center}
\includegraphics[width=4in]{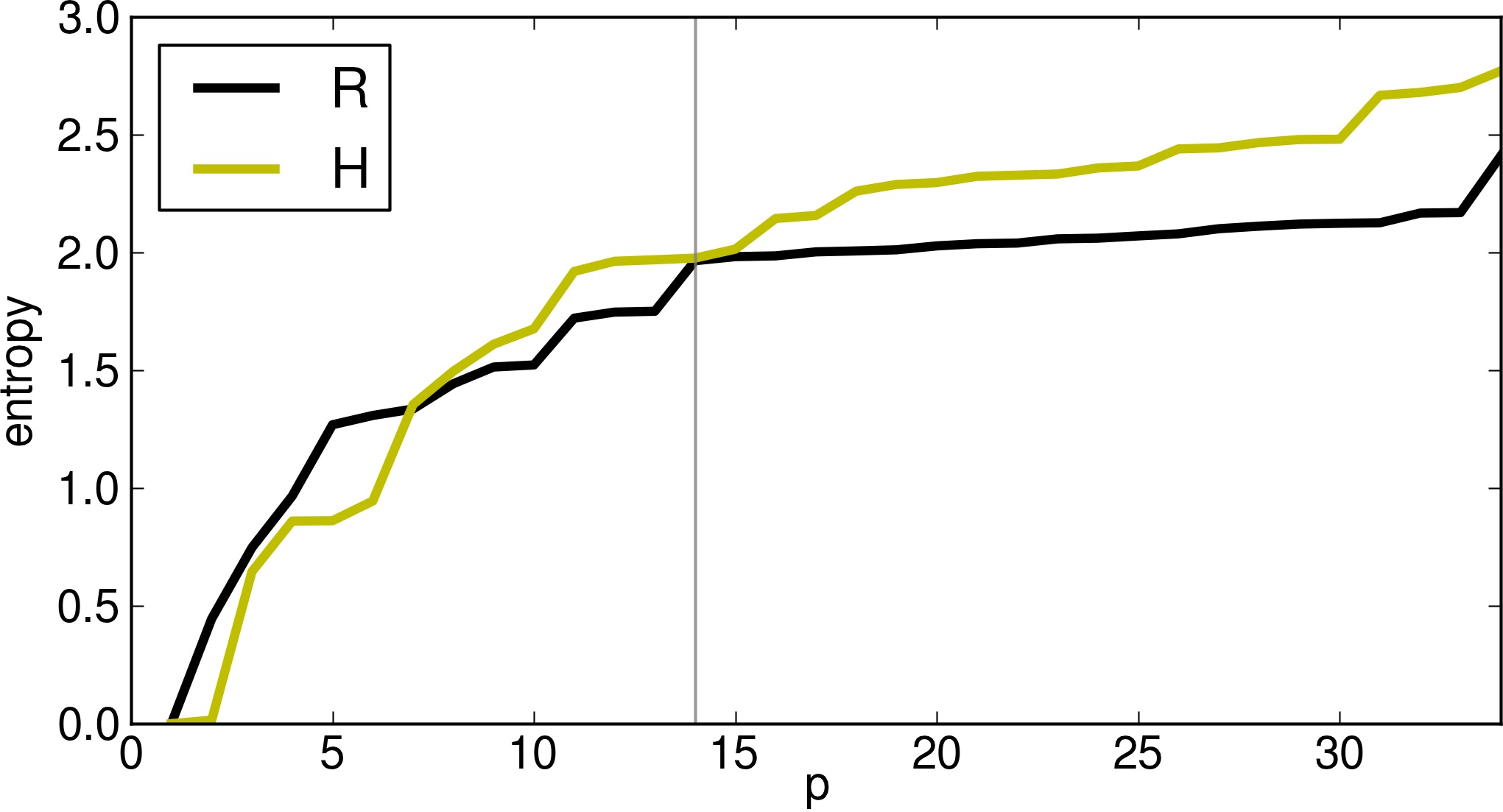}
\end{center}
\caption{
{\bf Looking for optimal $p$ in clustering.}  Clustering entropy for Hamming and Rohlin
distance at different $p$ values for H3N2. The plateau, in Rohlin, suggests an
optimal and stable result for the clustering.
}
\label{fig01}
\end{figure}

\newpage
\begin{figure}[!ht]
\begin{center}
\includegraphics[width=4in]{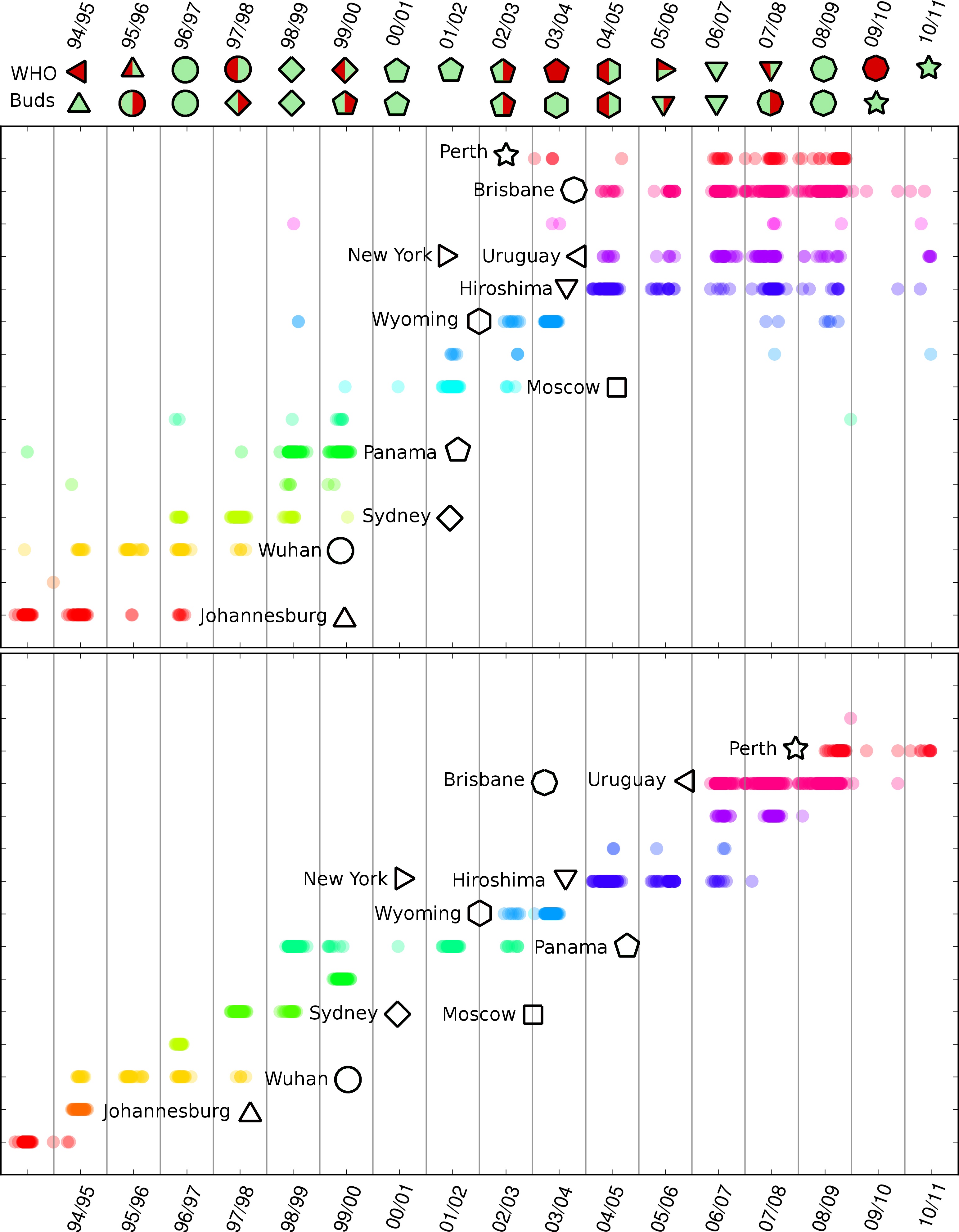}
\end{center}
\caption{
{\bf Clusters  and time evolution for H3N2.}  (upper part) Rohlin clusters time evolution for H3N2.
(lower part) Hamming clusters time evolution for H3N2. The reference WHO
sequences are shown by the corresponding symbols and names, and the details are indicated in Table 1. In
the upper part, we indicate the vaccine choice according to the WHO indication
(up) and to buds criterion (lower). Green and red colors indicate right and
wrong choice  with respect to the corresponding analysis on the real
circulating strain. A double color is used when more than one strain circulated in that year and the corresponding prediction agrees with one of the circulating strain.}
\label{fig02}
\end{figure}

\newpage
\begin{figure}[!ht]
\begin{center}
\includegraphics[width=4in]{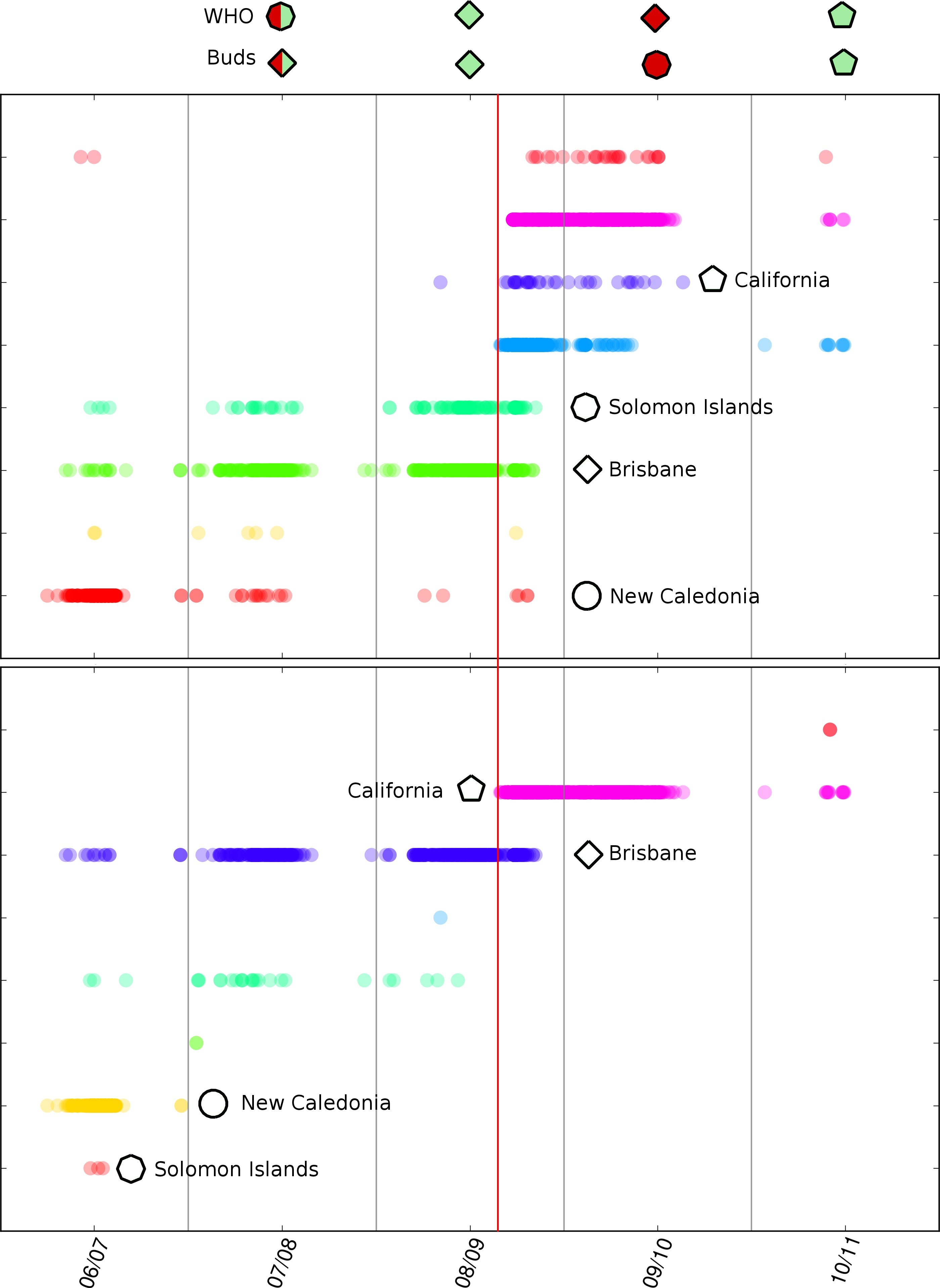}
\end{center}
\caption{
{\bf Clusters and time evolution for H1N1.} (upper part) Rohlin clusters time evolution for H1N1.
(lower part) Hamming clusters time evolution for H1N1. The reference WHO
sequences are shown by the corresponding symbols, as indicated in Table 2. In
the upper part, we indicate the vaccine choice according to the WHO indication
(up) and to buds criterion (lower). Green and red colors indicate right and
wrong choice. In this case, some of the symbols does not correspond to a
specific HI test, so they are indicated by a star and a pentagon. Notice the
onset of the pandemic virus and the failure of the bud criterion after that
line.
}
\label{fig03}
\end{figure}

\newpage
\begin{figure}[!ht]
\begin{center}
\includegraphics[width=4in]{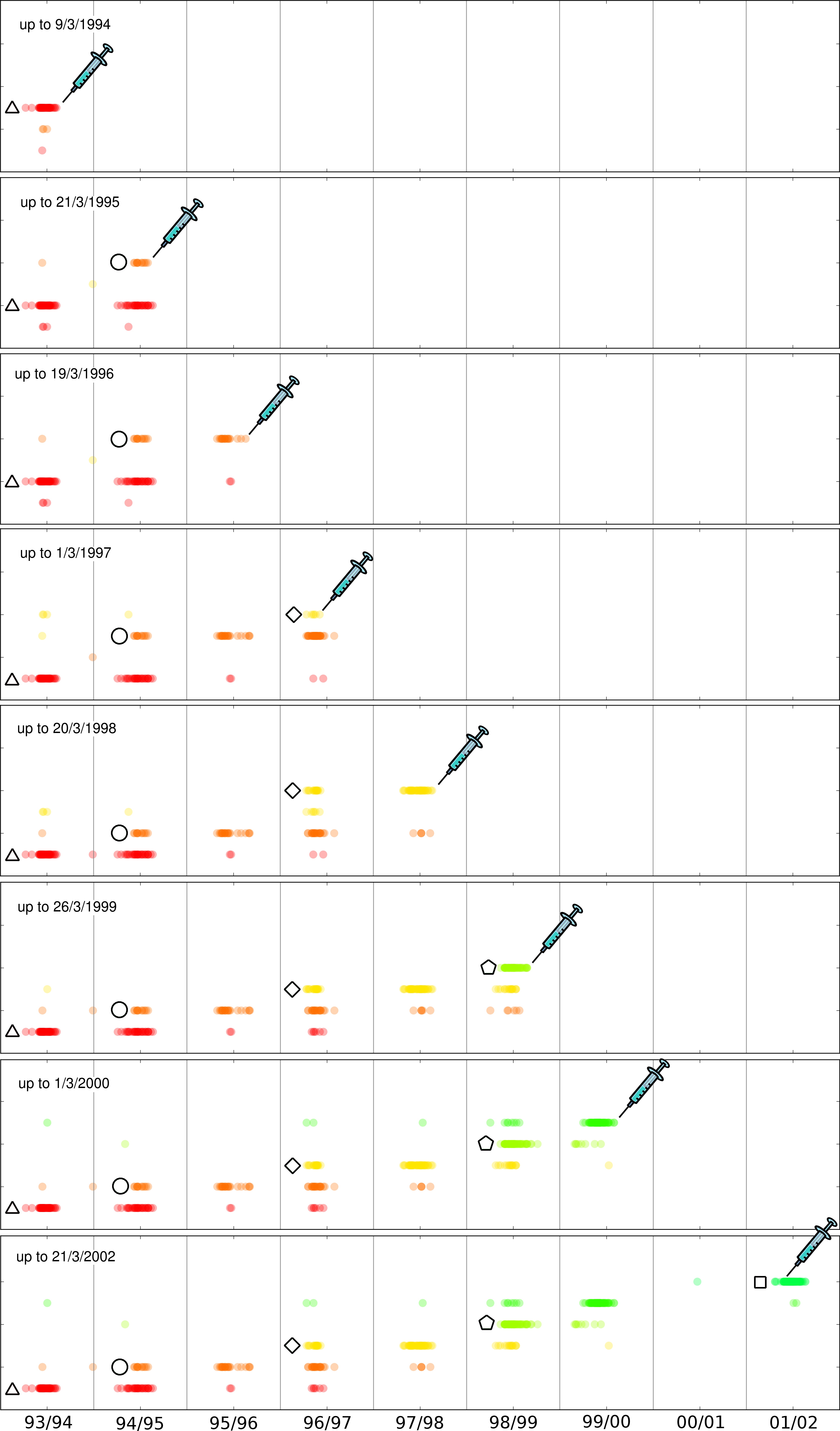}
\end{center}
\caption{
{\bf Changing the time window.} Rohlin analysis during years. Clusters structure is robust
under a sampling increase,  and buds appeared during seasons correctly reveal
the future circulating strains, as indicated by the syringe.
}
\label{fig04}
\end{figure}

\newpage
\begin{figure}[!ht]
\begin{center}
\includegraphics[width=4in]{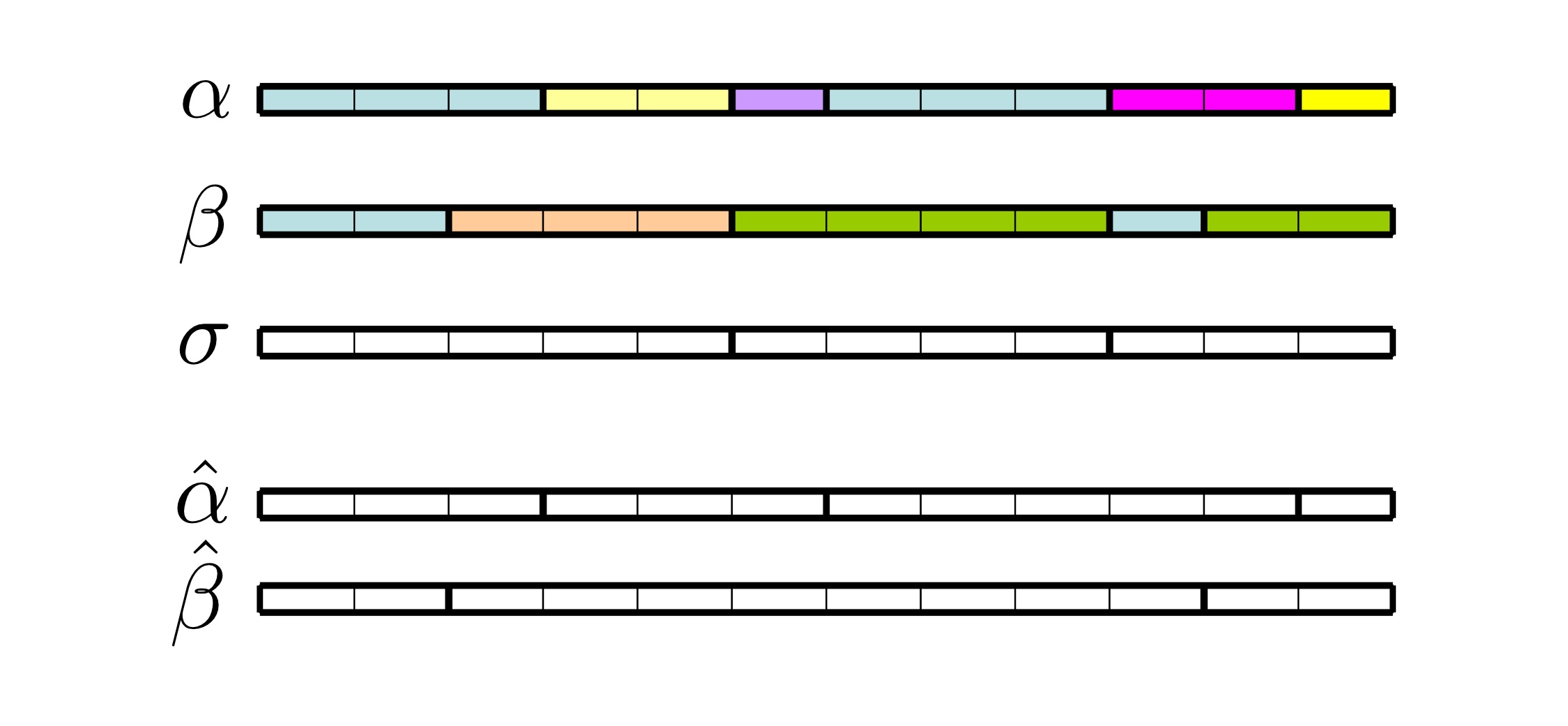}
\end{center}
\caption{
{\bf Reduction.} Two partitions $\alpha$ and $\beta$ (derived from sequences as given in the text) and their maximal common factor $\sigma$. Thick vertical lines individuate the atomic segments. Below, the reduced partitions $\widehat\alpha$ and $\widehat\beta$. Colors in $\alpha$ and $\beta$  remind the source configurations. There are no colors in $\sigma$, $\widehat\alpha$ and $\widehat\beta$ because these partitions do not originate from sequences of symbols but are directly defined in the partition space.
}
\label{fig05}
\end{figure}

\newpage
\section*{Tables}

\begin{table}[h!]
\centering
\caption{Symbols legend for Fig. 2:}
\begin{tabular}{|c|l|l|}
\hline
$\APLup$ &
A/Johannesburg/33/1994 &
AY661180\cr
\hline
\Circle &
A/Wuhan/359/1995 &
AY661190\cr
\hline
$\Diamond$ &
A/Sydney/5/1997 &
EF566075\cr
\hline
\pentagon &
A/Panama/2007/1999 &
DQ508865\cr
\hline
\Square &
A/Moscow/10/1999 &
DQ487341\cr
\hline
\hexagon &
A/Wyoming/03/2003 &
CY034108\cr
\hline
$\rhd$ &
A/New York/55/2004 &
CY033638\cr
\hline
$\APLdown$&
A/Hiroshima/52/2005 &
EU283414\cr
\hline
\octagon &
A/Brisbane/10/2007 &
CY035022\cr
\hline
$\lhd$&
A/Uruguay/716/2007 &
EU716426\cr
\hline
\ding{73} &
A/Perth/16/2009 &
GQ293081\cr
\hline
\end{tabular}
\end{table}

\begin{table}[h!]
\centering
\caption{Symbols legend for Fig. 3:}
\begin{tabular}{|c|l|l|}
\hline
\Circle &
A/New Caledonia/20/1999 &
CY033622\cr
\hline
\hexagon &
A/Solomon Islands/3/2006 &
EU124177\cr
\hline
$\Diamond$ &
A/Brisbane/59/2007 &
CY058487\cr
\hline
\pentagon &
A/California/07/2009 &
FJ969540\cr
\hline
\end{tabular}
\end{table}

\newpage
\section*{Supporting Information Legends} 

\begin{figure}[!ht]
\begin{center}
\includegraphics[width=4in]{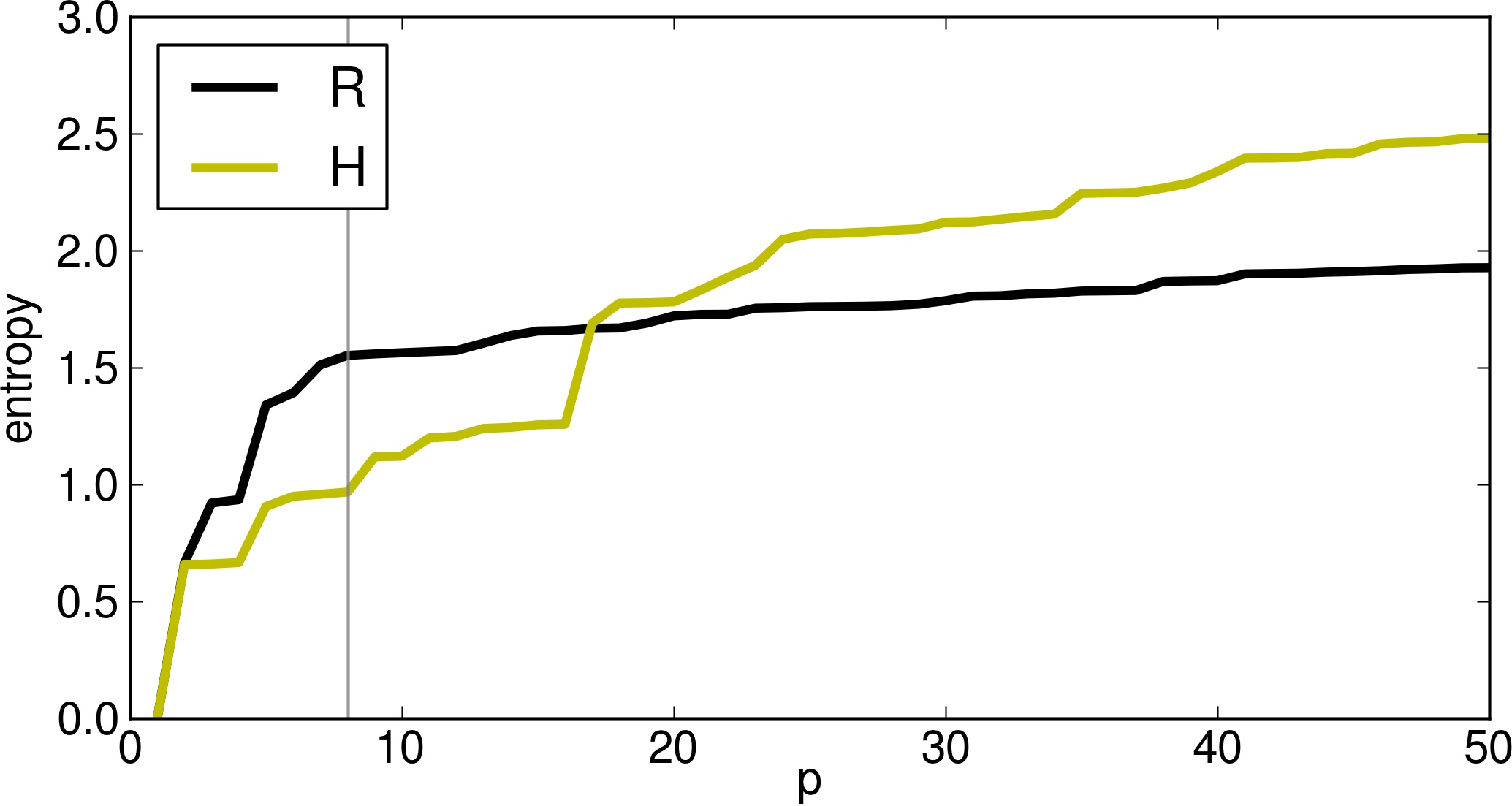}
\end{center}
\caption*{
{\bf Figure S1: Looking for optimal $p$ in clustering  for H1N1.} Clustering  entropy for Rohlin and Hamming at different
$p$ values for influenza A H1N1. The long plateau, in Rohlin, suggests a stable
and well defined value for the optimal $p$. Notice that Hamming is growing.
}
\label{figS1}
\end{figure}

\newpage
\begin{figure}[!ht]
\begin{center}
\includegraphics[width=4in]{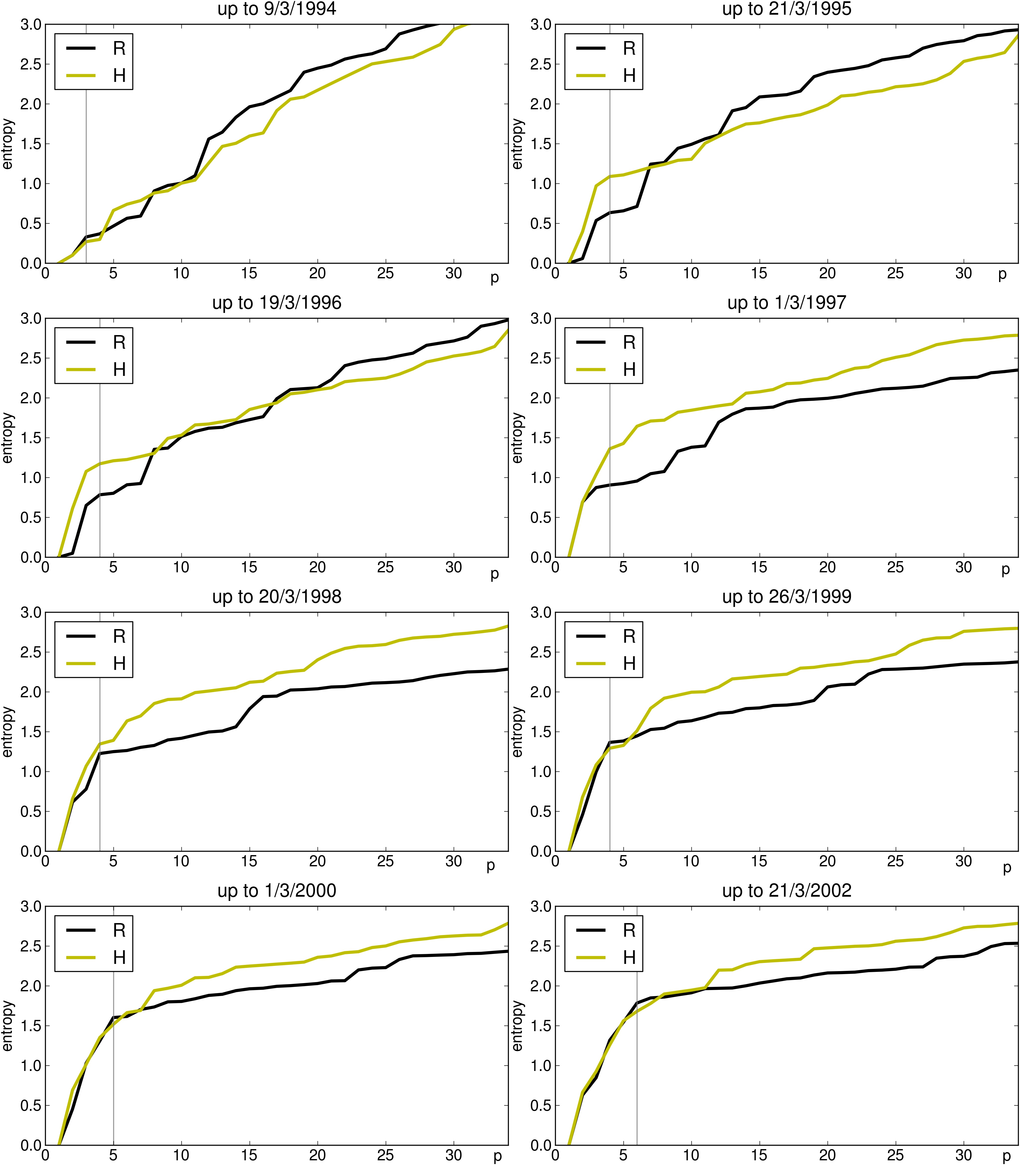}
\end{center}
\caption*{
{\bf Figure S2: Looking for optimal $p$ in clustering  for H3N2 in the restricted time window.} Clustering  entropy for Rohlin and Hamming at different
$p$ values for influenza A H3N2, as obtained by considering only the sequences up to the end 
of the winter season of the year indicated in the plot. In each time window, the long plateau, in Rohlin, suggests a stable
and well defined value for the optimal $p$. This figure is in correspondence with Fig. 4 of the main text.
}
\label{figS2}
\end{figure}

\newpage
\begin{figure}[!ht]
\begin{center}
\includegraphics[width=4in]{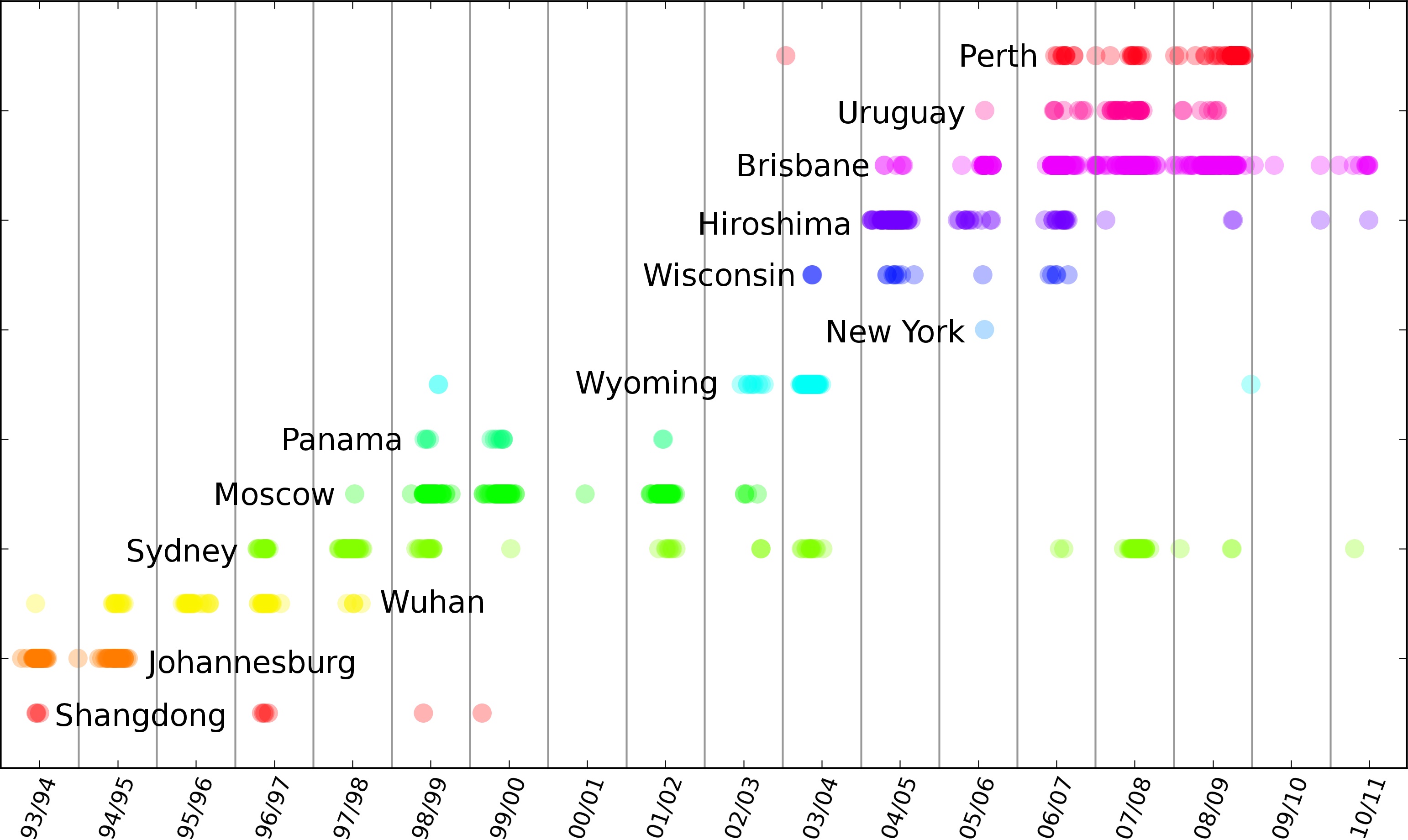}
\end{center}
\caption*{
{\bf Figure S3: Reverse analysis for Rohlin clusters.} Sequences of minimum distance with the
corresponding WHO reference sequences, during years. The great similarity with Fig. 2
shows a strong consistency between Rohlin and HI analysis.
}
\label{figS3}
\end{figure}

\begin{figure}[!ht]
\begin{center}
\includegraphics[width=4in]{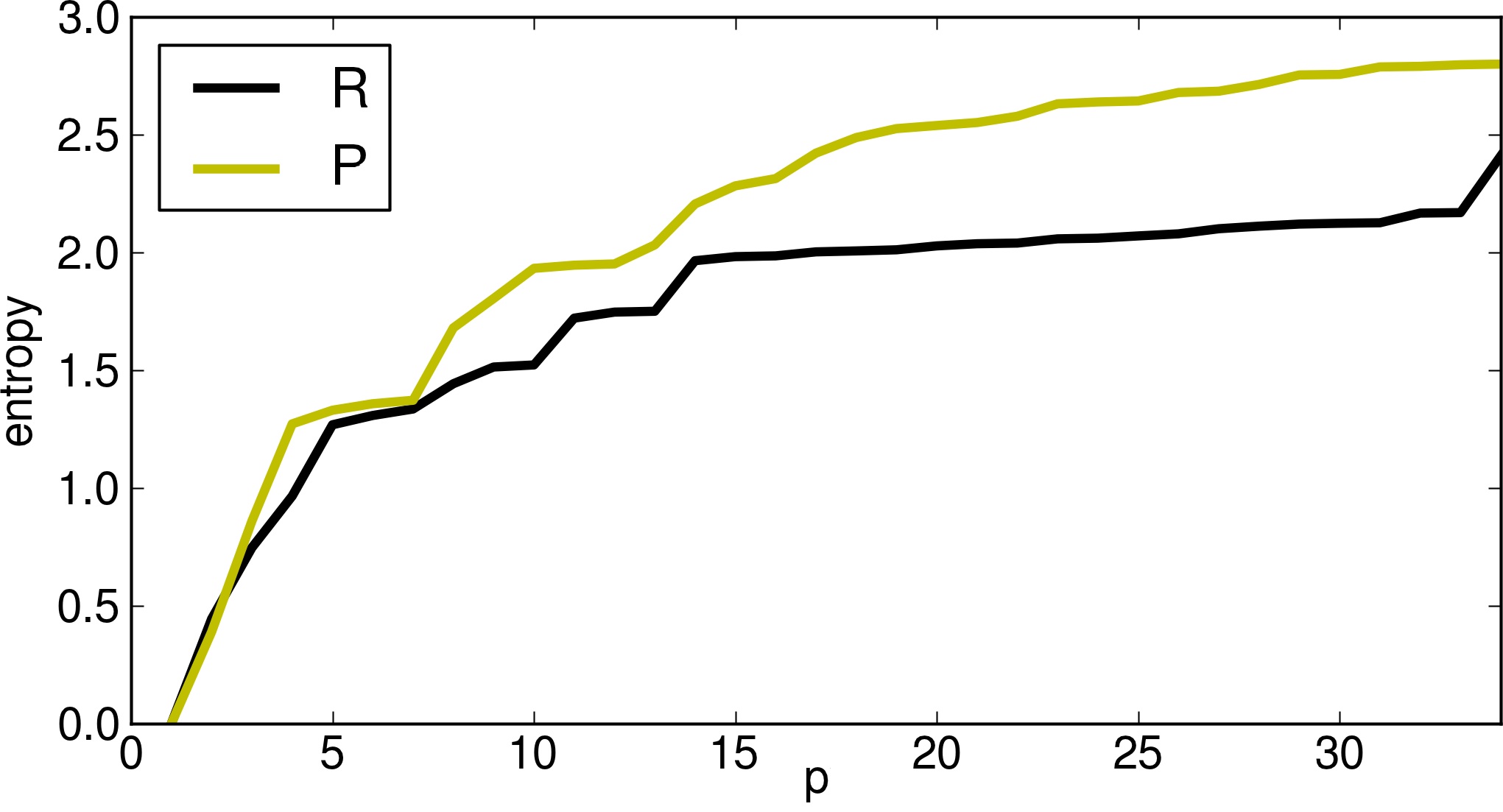}
\end{center}
\caption*{
{\bf Figure S4: Clustering on random permutations.} Effect of random permutation of symbols on the entropy of
the clustering, as a function of $p$.  R indicates the entropy of clustering with the Rohlin distance and P stands for the entropy of clustering in the sample, obtained under a random permutation of symbols in each
sequence. 
}
\label{figS4}
\end{figure}

\end{document}